%% file: main_icml.tex
\documentclass{article}

\usepackage{microtype}
\usepackage{graphicx}
\usepackage{booktabs} 
\input{math_commands.tex}

\usepackage{subcaption}
\usepackage{adjustbox}

\usepackage{hyperref}

\usepackage[accepted]{icml2024}

\usepackage{amsmath}
\usepackage{amssymb}
\usepackage{mathtools}
\usepackage{amsthm}
\usepackage{tikzducks}
\usepackage{glossaries}

\usepackage[capitalize,noabbrev]{cleveref}

\usepackage{comment}
\usepackage{todonotes}

\glsdisablehyper
\loadglsentries{glossary}

\theoremstyle{plain}

\theoremstyle{definition}

\theoremstyle{remark}

\title{Equation identification for fluid flows via physics-informed neural networks}

\icmltitlerunning{Equation identification for fluid flows via physics-informed neural networks}

\begin{document}

\twocolumn[
\icmltitle{Equation identification for fluid flows via physics-informed neural networks}




\begin{icmlauthorlist}
\icmlauthor{Alexander New}{xxx}
\icmlauthor{Marisel Villafañe-Delgado}{xxx}
\icmlauthor{Charles Shugert}{xxx}
\end{icmlauthorlist}

\icmlaffiliation{xxx}{Johns Hopkins University Applied Physics Laboratory, 11100 Johns Hopkins Rd, Laurel, MD 20723, USA}

\icmlcorrespondingauthor{Alexander New}{alex.new@jhuapl.edu}

\icmlkeywords{Machine Learning, ICML}

\vskip 0.3in
]



\printAffiliationsAndNotice{}  

\begin{abstract}
\Gls{SciML} methods such as \glspl{PINN} are used to estimate parameters of interest from governing equations and small quantities of data. However, there has been little work in assessing how well \glspl{PINN} perform for inverse problems across wide ranges of governing equations across the mathematical sciences. We present a new and challenging benchmark problem for inverse \glspl{PINN} based on a  parametric sweep of the 2D Burgers' equation with rotational flow. We show that a novel strategy that alternates between first- and second-order optimization proves superior to typical first-order strategies for estimating parameters. In addition, we propose a novel data-driven method to characterize \gls{PINN} effectiveness in the inverse setting. \Glspl{PINN}' physics-informed regularization enables them to leverage small quantities of data more efficiently than the data-driven baseline. However, both \glspl{PINN} and the baseline can fail to recover parameters for highly inviscid flows, motivating the need for further development of \gls{PINN} methods.
\end{abstract}

\glsresetall

\section{Introduction}\label{sec:introduction}

\Glspl{PINN}~\cite{raissi2019physics} have emerged as a practical tool for the solution of \glspl{PDE} in many scientific applications modeling complex systems. \glspl{PINN} overcome some of the limitations of classical \gls{PDE} solvers, improving on the expensive computational requirements of generating grids and developing bespoke solving schemes~\cite{Ma2021photons,Raissi2020fluid,Ning2023elastic,Haghiat2021solid,Alber2019biology,Cai2021heat,Hao2023pinnacle}. 



\Glspl{PINN} can be used to solve both forward and inverse problems. For the \gls{PINN} forward problem, a number of challenges and pathologies have been identified that arise when training \glspl{PINN}, and mitigation methods have been suggested~\cite{wang2021understanding,wang2022ntk,Wang2022causality,krishnapriyan2021characterizing,New2023pinns,Basir2022equality,mclenny2020adaptive,Daw2022mitigating,Maddu2022dirichlet,Lu2021exact,Sukumar2022distancefunctions}.

In contrast, the inverse problem has seen less systematic study, although some recent benchmark have included them~\cite{Hao2023pinnacle}. Inverse problems are of particular interest because, in practice, the \glspl{PDE} modeling physical phenomena are often partially specified. Strategies to mitigate some of the challenges specific to the \gls{PINN} inverse problem include improved optimization~\cite{yu2022gradient}, new architectures ~\cite{aliakbari2023ensemble}, and temporal methods for time-dependent \glspl{PDE}~\cite{mattey2022novel}.

One challenge arising in the solution of inverse problems with \glspl{PINN} is the limited range of evaluated \glspl{PDE}. Studies typically consider a single \gls{PDE} instance and do not vary parameters or other conditions. However, these parameters or constants (e.g. Reynolds number in fluid flow) that \glspl{PDE} include may vary across application areas. 

Additionally, the estimation of \gls{PDE} parameters with \glspl{PINN} presents particular challenges not encountered in the forward setting. While the forward setting requires fitting \gls{NN} parameters, in the inverse problem, the challenge is to fit both the \gls{NN} and the \gls{PDE} parameters, which differ significantly in their dimensionality. This results in limitations for methods such as Adam~\cite{Kingma2014adam} in the solution of inverse problems with \glspl{PINN}.

Our contributions to the inverse problem for \glspl{PINN} are three-fold. First, we propose a novel strategy for estimating \gls{PDE} parameters using \glspl{PINN} that alternates between using \gls{SGD} to update \gls{NN} weights and using Newton's method to estimate \gls{PDE} parameters. Second, we introduce a 2D Burgers' equation benchmark problem with varying parameter coefficients across $10$ solutions, including highly viscous and inviscid flows. Third, we propose an approach for estimating the benefit of using physics-informed regularization in \gls{PDE} estimation problems. We show that the \glspl{PINN}' use of physics-based regularization enhances its effectiveness in parameter estimation problems, although \glspl{PINN} can still struggle to simultaneously minimize loss criteria involving \glspl{PDE} and data.

The rest of this paper is organized as follows: In~\cref{sec:problem,sec:burgers,sec:related_work}, we introduce the \gls{PDE} inverse problem and explain how it applies to the 2D Burgers' equation, and in~\cref{sec:pinns}, we present an overview of \glspl{PINN}. \cref{sec:estimation}, ~\cref{sec:data_driven}, and ~\cref{sec:data} describe our main contributions in this work, highlighting the unique challenges that occur in the \gls{PINN} inverse problem. Finally, in~\cref{sec:implementation,sec:comparison,sec:analysis}, we evaluate methods on the 2D Burgers' equation.




\section{Methods}\label{sec:methods}

\subsection{Problem formulation}\label{sec:problem}

We consider a function $\mathbf{u}:\Omega \to \mathbb{R}^m$ defined on $\Omega \subseteq \mathbb{R}^n$. The function $\mathbf{u}$ satisfies a set of differential operators $N_i(\cdot;\phi)$ for $i\in I_N$. The domain $\Omega$ has $r$ boundaries $\Gamma_j$, $j=1,\hdots,r$ (including, possibly, the beginning of a temporal dimension). The function $\mathbf{u}$ satisfies \gls{BC} or \gls{IC} operators $B_j(\cdot;\phi)$ on boundaries $\Gamma_j$, for a set of indices $j \in I_b$. Here, $\phi$ is one or more scalar parameters defining the operators $N_i$ and $B_j$. Then the \gls{PDE} may be written as:
\begin{eqnarray}
N_i(\mathbf{u}; \phi)(\mathbf{x}) &=& 0\,\,\,\,\,\mathbf{x} \in \Omega,\,\,\,\,i \in I_N\nonumber\\
B_j(\mathbf{u}; \phi)(\mathbf{x}) &=& 0\,\,\,\,\,\mathbf{x} \in \Gamma_j,\,\,j \in I_B. 
\label{eq:pde}
\end{eqnarray}

The type of \gls{BC} can include Dirichlet \glspl{BC} ($B_j(\mathbf{u}; \phi)(\mathbf{x}) = \mathbf{u}(\mathbf{x}) - \mathbf{f}_j(\mathbf{x}; \phi)$ for a function $\mathbf{f}_j$), Neumann \glspl{BC} ($B_j(\mathbf{u}; \phi)(\mathbf{x}) = (\hat{\mathbf{n}}\cdot\nabla \mathbf{u})(\mathbf{x}) - \mathbf{g}_j(\mathbf{x}; \phi)$ for a function $\mathbf{g}_j$), periodic \glspl{BC}, or another type. The differential operators $N_i$ might be linear or nonlinear. A \gls{BC} could be parameterized by the speed of an inflow \gls{BC}~\cite{Collins2023rapid}, or the domain $\Omega$ could be parameterized by the size and shape of a defect in its interior~\cite{zhang2023sciml}.


Given  samples $\mathcal{D} = \{(\mathbf{u}_n, \mathbf{x}_n)\}_{n=1}^N$ from the solution, the inverse problem's goal is to (i) obtain a function that can fill in $\mathbf{u}$ for the remainder of the domain, and (ii) identify values for $\phi$ consistent with the training data and \glspl{BC}.


\subsection{The 2D Burgers' equation}\label{sec:burgers}


\begin{figure}[h]
    \centering
    \includegraphics[width=\linewidth]{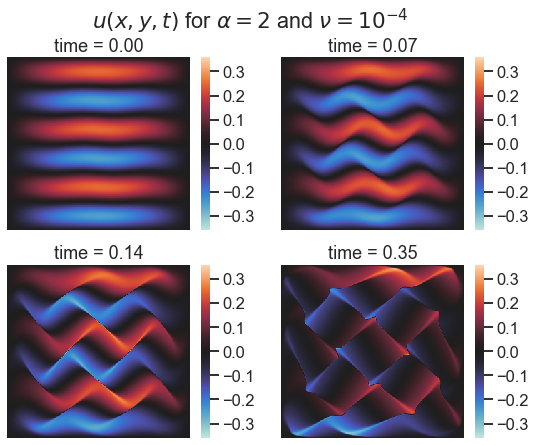}
    \caption{We show a few temporal snapshots of the $u$-component of a Burgers' equation solution. At $t=0$, the solution exhibits variation only in the $y$-direction; as time progresses (e.g., for $t=0.14$ and $t=0.35$), rotational flow develops.}
    \label{fig:solution_visualization}
\end{figure}

In this work, we focus on the sourceless time-varying Burgers' equation. It has two solution components ($\mathbf{u} = (u, v)$), two spatial inputs and one temporal input ($\mathbf{x} = (x, y, t)$) defined on the domain $[0, 1]\times[0,1]\times[0,0.5]$:
\begin{eqnarray}
\rvu_t + \alpha\,\rvu \cdot \nabla \rvu - \nu \Delta \rvu &=& 0\label{eq:burgers}\\
u(x, y, 0) &=& \sin(6 \pi y) x (1 - x)\label{eq:ic}\\
v(x, y, 0) &=& -\sin(6 \pi x) y (1 - y)\nonumber\\
\rvu(x, 0, t) = \rvu(x, 1, t) &=& 0\label{eq:uv_bc}\\
\rvu(0, y, t) = \rvu(1, y, t) &=& 0.\nonumber
\end{eqnarray}
The parameters $\phi = \{\alpha, \nu\}$ characterize the convection $\alpha$ and diffusion $\nu$ components of the solution. When $\alpha=0$, eq.~\ref{eq:burgers} reduces to a series of uncoupled heat equations. When $\nu = 0$, eq.~\ref{eq:burgers} reduces to an inviscid Burgers' equation.

Eq.~\ref{eq:burgers} can be reparameterized into the usual Burgers' equation by rescaling the domain with the variable transformation $\mathbf{x} \rightarrow \alpha\,\mathbf{x}$: 
\begin{eqnarray}
    \rvu_t + \alpha\,\rvu \cdot \nabla \rvu - \nu \Delta \rvu
    &\rightarrow& \rvu_t + \rvu \cdot \nabla \rvu - \frac{\nu}{\alpha^2} \Delta \rvu\nonumber. 
\end{eqnarray}

After  this reparameterization, the Cole-Hopf transformation~\cite{Cole1951OnAQ,Hopf1950ThePD} can be used to analytically solve the Burgers' equation, so long as the solution does not admit irrotational flow (i.e., if $\nabla \times \mathbf{u} = 0$). Thus, we choose \glspl{IC} that vanish at the boundaries but have nonzero curl. In this way, the data we generate is an example of the simplest nontrivial (rotational) Burgers' flow. 


\subsection{Existing uses of PINNs for Burgers' equation}\label{sec:related_work}

To the best of our knowledge, there has not been a previous study of the 2D Burgers' equation that uses two solution components and varies its parameters. In general, the 1D Burgers' equation is a common benchmark problem for forward \glspl{PINN} ~\citet{ren2022phycrnet, carniellopinn, mathias2022augmenting, rosofsky2023applications}, but there has been less work studying the 2D Burgers' equation, especially in the inverse problem setting for varying parameters \citet{Alkhadhr2021burgers, xu2023discovery}. 



In the forward setting,~\citet{ren2022phycrnet} proposed a convolutional-recurrent architecture and evaluated its performance in an extrapolation setting on the 2D Burgers' problem with two components and fixed viscosity. \citet{carniellopinn} proposed a \gls{PINN} to solve the scalar 2D inviscid Burgers' with data from a Riemann problem, which exhibits discontinuities and refraction and shock waves. \citet{mathias2022augmenting} proposed a data augmentation strategy and evaluated its performance in the 2D Burgers' forward problem with two components. \citet{kim2022fast} proposed a nonlinear manifold \gls{ROM} for estimating the two components of the viscous Burgers' with large Reynolds numbers. \citet{rosofsky2023applications} evaluated various settings of the Burgers' equation in 1D and 2D: 2D scalar, 2D inviscid, 2D vector Burgers'  using a physics-informed neural operator.  In the inverse problem setting, \citet{Alkhadhr2021burgers} proposed PINNs to solve the 1D and 2D Burgers' in the forward and inverse problems, however, their solution has a single component, and they only consider a single set of \gls{PDE} parameters. \citet{xu2023discovery} also solved the 2D Burgers' inverse problem for equation discovery, but with one component. 


\subsection{Physics-informed neural networks}\label{sec:pinns}

In the inverse problem \gls{PINN} approach~\cite{raissi2019physics}, a neural network $\mathbf{u}_\theta$ parameterized by weights $\theta$ is trained to satisfy both the \gls{PDE} and the data:
\begin{eqnarray}
\theta^*, \phi^* &=& \arg\min_\theta L(\theta; \phi),
\label{eq:minimization}
\end{eqnarray}
where
\begin{eqnarray}
L(\theta; \phi) &=& L_{PDE}(\theta; \phi) + L_{BC}(\theta; \phi)\nonumber \\
&&+ L_{Data}(\theta)\,\,\,\,\,\,\,\,\,\,\,\,\,\,\,\,\,\,\,\,\,\,\,\,\,\,\,\,\,\,\,\,\,\,\,\,\,\,\,\,\,\,\,\,\,\,\,\,\,\,\,\,\,\,\,\,\,\,\,\,\,\,\,\,\,\label{eq:pinn_loss}\\
L_{PDE}(\theta; \phi) &=& \sum_{i \in I_N} \lambda_i \sum_{\mathbf{x}_\Omega \in \Omega} ||N_i(\mathbf{u}_\theta; \phi)(\mathbf{x}_\Omega)||^2\label{eq:residual}\\
L_{BC}(\theta) &=& \sum_{j \in I_B}\lambda_j\sum_{\mathbf{x}_j \in \Gamma_j}||B_j(\mathbf{u}_\theta;\phi)(\mathbf{x}_j)||^2\label{eq:bc}\\
L_{Data}(\theta) &=& \lambda_{Data} \sum_{n=1}^N ||\mathbf{u}_n - \mathbf{u}_\theta(\mathbf{x}_n)||^2.
\label{eq:data}
\end{eqnarray}
Here, $\lambda_i$, $\lambda_j$, and $\lambda_{Data}$ are nonnegative weights for each component of the loss function, $\{\mathbf{x}_\Omega\}$ is a set of points sampled from the domain $\Omega$, and $\{\mathbf{x}_j\}$ is a set of points sampled from each boundary $\Gamma_j$. The set of points can either be chosen at the start of training (via random sampling or a chosen discretization), or it can change or grow during training~\cite{Daw2022mitigating,lu2021deepxde}.

A noted challenge in training \glspl{PINN} is that their loss function can contain many components that must all be minimized for the problem to be solved. To reduce the complexity of the \gls{PINN} training process and focus on the underlying inverse problem, we use an \gls{NN} that vanishes by construction at the domain's boundaries:
\begin{eqnarray}
    u_\theta(x, y, t) &=& x(1-x)y(1-y)\mathrm{NN}_u(x, y, t; \theta)\nonumber\\
    v_\theta(x, y, t) &=& x(1-x)y(1-y)\mathrm{NN}_v(x, y, t; \theta).\,\,\,\,\,\,\label{eq:exact}
\end{eqnarray}
Here $\mathrm{NN}$ is a \gls{MLP} or \gls{EDMLP}~\cite{wang2021understanding} with two output nodes ($\mathrm{NN}_u$ and $\mathrm{NN}_v$). Similar strategies use periodic~\glspl{MLP}~\cite{Dong2021periodic,Wang2022causality} when solving \glspl{PDE} on periodic domains.

\subsection{Estimating PDE parameters using neural networks}\label{sec:estimation}


In the forward problem, \glspl{PINN} are often trained with a combination of first-order optimization methods like \gls{SGD} combined with Adam~\cite{Kingma2014adam} and approximate second-order methods like L-BFGS~\cite{Liu1989lbfgs}. However, estimating \gls{PDE} parameters from a partially-specified set of governing equations and some amount of data differs in several key respects from the typical training done to fit \gls{NN} parameters.

In particular, $\phi$'s dimensionality is much lower than $\theta$'s. Here, $\phi$ has only two components, while the typical \gls{NN} may have between $10^5$ and $10^7$ weights. Furthermore, the scale of the components of $\phi$ has a physical meaning based on the governing equations and domain size. These factors combine to make techniques like Adam~\cite{Kingma2014adam} or MultiAdam~\cite{Yao2023MultiAdam} inefficient for estimating \gls{PDE} parameters using \glspl{PINN}.
For example, misspecified step sizes for $\phi$ can cause its iterates to diverge.

In this work, we use Newton's method to estimate $\phi$ but rely on Adam for fitting $\theta$. To motivate this, observe that, for the Burgers' equation, each governing equation (eq.~\ref{eq:burgers}) is a linear function of its parameters $\phi = \{\alpha, \nu\}$. This makes the \gls{PINN} loss (eq.~\ref{eq:pinn_loss}) convex and quadratic in $\phi$, meaning that, for fixed $\theta$, a single Newton step yields the optimal $\phi$.\footnote{It is not generally the case that the \gls{PINN} learning problem will be linear in its parameters.} To the best of our knowledge, this observation has not been widely discussed in the inverse \gls{PINN} literature.



\subsection{Data-driven \gls{PDE} parameter estimation}\label{sec:data_driven}

\begin{algorithm}
    \caption{Estimating \gls{PDE} parameters using a data-driven \gls{NN}}
    \begin{algorithmic}[1]
        \REQUIRE Sample of \gls{PDE} solutions $\{(\mathbf{u}_n, \mathbf{x}_n)\}_n^N$
        \REQUIRE Differential operators $N_i$ and boundary condition operators $B_j$
        \REQUIRE Neural network architecture $\mathbf{u}_\theta$\\
        \STATE Minimize $\hat{\theta} = \arg\min_\theta L_{BC}(\theta) + L_{Data}(\theta)$ with \gls{SGD}
        \STATE Minimize $\hat{\phi} = \arg\min_\phi L_{PDE}(\hat{\theta}; \phi)$ with Newton's method
        \STATE Return \gls{PDE} parameter estimates $\hat{\phi}$
    \end{algorithmic}
    \label{alg:estimation}
\end{algorithm}

An understudied challenge in the use of \glspl{PINN} is the question of how much labeled data is needed to reconstruct the full solution and accurately estimate the \gls{PDE} parameters. For example, the Poisson and diffusion inverse problems studied in~\cite{Hao2023pinnacle} use a fixed labeled dataset size of $2500$ points. Especially for problems involving nonlinear \glspl{PDE}, existing theory may not be sufficient to give conditions for a problem to be identifiable.


Thus, here we propose a novel baseline for assessing the data efficiency benefits gained by using \glspl{PINN} for inverse problems. It is summarized in~\cref{alg:estimation}. For a fixed sample of labeled data, we train a \gls{PINN} to minimize the sum of the \gls{BC} loss (eq.~\ref{eq:bc}) and the data loss (eq.~\ref{eq:data}). We then use Newton's method to estimate the optimal \gls{PDE} parameters that minimize the \gls{PDE} residual (eq.~\ref{eq:residual}) while holding the \gls{NN} weights constant. \Cref{alg:estimation} relies on the \gls{BC} loss $L_{BC}$  being independent of the parameters $\phi$. If this were not the case, then the first step would train the \gls{NN} using only $L_{data}$.

For a sufficiently large quantity of labeled data and a sufficiently expressive \gls{NN}, we expect the \gls{NN} to interpolate the full discretized solution. If the discretization error is not too high, this should enable the parameters $\phi$ to be accurately recovered. The benefit of a \gls{PINN} may be observed in the data quantity regime where a physics-unaware \gls{NN} cannot fit the entire solution, but a \gls{PINN} can.




\subsection{Data generation}\label{sec:data}

\begin{table}[h]
    \centering
    \begin{tabular}{c|c}
    $\nu$       &   $\alpha$\\\hline
    $10^{-4}$   &   $1.0, 1.5, 2.0$\\
    $10^{-3}$   &   $1.0, 1.5, 2.0$\\
    $10^{-2}$   &   $0.5, 1.0, 1.5, 2.0$
    \end{tabular}
    \caption{The parameter values of the Burgers' equation (eq.~\ref{eq:burgers}) used to generate ground truth solutions.}
    \label{tab:burgers_parameters}
\end{table}

Solutions to the Burgers' equation were obtained by using FiPy: A Finite Volume PDE Solver~\cite{FiPy}. We use a $256\times256$ uniform discretization of $[0,1]\times[0,1]$ for the spatial domain. There are $72$ timesteps that cover $t=0$ to $t=0.499$, which means that solutions have $256\cdot256\cdot72= 4{,}718{,}592$ points total per simulation. 

The diffusive term in the Burgers' equation was handled 
using an implicit scheme. To handle the nonlinear convective term, the Burgers' Equation was solved in its conservative form.
The convective $\mathbf{u}\cdot\nabla\mathbf{u}$ terms were solved using an implicit discretization scheme, and the $\Delta \mathbf{u}$ terms were handled using a Power-Law discretization scheme~\cite{versteeg2007introduction}. At each time step, the \gls{PDE} used a linear LU solver with an Algebraic Multigrid Preconditioner~\cite{pyamg2023}.

Using the parameter values in~\cref{tab:burgers_parameters}, we generate $10$ ground truth solutions to the Burgers' equation. We view $\alpha/\nu$ as an effective-order parameter for the system's Reynolds number. By varying $\nu$ logarithmically, we study how effective the \glspl{PINN} are in both highly viscous and highly inviscid regimes. In contrast, we vary $\alpha$ linearly to study how sensitive \glspl{PINN} are at learning parameters that have smaller effects on the governing fluid equations.

\section{Results}\label{sec:results}

\subsection{Implementation and evaluation procedure}\label{sec:implementation}

For each \gls{PDE} parameter value and amount of training data ($2048$, $8192$, $32768$ points), we train \glspl{PINN} using both Adam and Newton's method for parameter value estimation. Similarly, for each \gls{PDE} parameter and amount of training data ($128$, $512$, $2048$, $8192$, $32768$, $131072$ points), we apply~\cref{alg:estimation} for data-driven parameter estimation.

We implement~\glspl{PINN} with \texttt{pinn-jax}~\cite{New2023pinns,New2024pinnjax}\footnote{\url{https://github.com/newalexander/pinn-jax}}, which uses \texttt{jax}~\cite{Bradbury2018jax}, \texttt{flax}~\cite{Heek2020flax}, and~\texttt{optax}~\cite{Babuschkin2020optax}. \gls{PDE} derivatives are calculated with the forward-mode \texttt{jacfwd} function~\cite{Baydin2017autodiff}. All computations use double precision. See~\Cref{sec:hyperparameters} for details on hyperparameters. For training \glspl{PINN}, we use Adam~\cite{Kingma2014adam}.


We initialize the convection coefficient $\alpha$ by sampling from $[0, 5]$ uniformly, and we initialize the diffusion coefficient $\nu$ by sampling from $[0, 0.5]$ uniformly. This choice is analogous to, in Bayesian methods for inverse problems, imposing a uniform prior on the unknown parameters (e.g.,~\cite{Christopher2018inverse,Doronina2020inverse}). 

For a ground truth solution $\mathbf{u}$, a trained solution $\mathbf{u}_\theta$, and a set of points $\mathbf{X}\subseteq\Omega$, we evaluate model predictions with the relative error $E_{rel}$:
\begin{eqnarray}
    E_{rel}(\mathbf{u}, \mathbf{u}_\theta; \mathbf{X}) = \frac{\left(\sum_{\mathbf{x}\in\mathbf{X}} ||\mathbf{u}(\mathbf{x}) - \mathbf{u}_\theta(\mathbf{x})||^2\right)^{1/2}}{\left(\sum_{\mathbf{x}\in\mathbf{X}}||\mathbf{u}(\mathbf{x})||^2\right)^{1/2}},
    \label{eq:rel_error}
\end{eqnarray}
where $\mathbf{X}$ is a uniform discretization of the domain and its boundaries. For reporting error in estimating $\alpha$, we use the scalar relative error $|\hat{\alpha} - \alpha| / \alpha$. Because $\nu$ takes values across different orders of magnitudes, we report relative error of logs: $|\log_{10}(\hat{\nu}+\epsilon) - \log_{10}(\nu+\epsilon)| / |\log_{10}(\nu+\epsilon)|$, where $\epsilon=10^{-8}$ prevents numerical overflow in the case that $\hat{\nu}=0$. We follow~\citet{Hao2023pinnacle} and consider a relative error that is $10\%$ or less to be sufficiently accurate for the estimation problem to be solved.

For \glspl{PINN}, our results are from the model checkpoint that attained the lowest loss (eq.~\ref{eq:pinn_loss}). For the data-driven estimation, our results are from the model checkpoint that attained the lowest error on the data (eq.~\ref{eq:data}).

\subsection{Methods assessment}\label{sec:comparison}

\begin{figure}[h]
    \centering
    \includegraphics[width=0.9\linewidth]{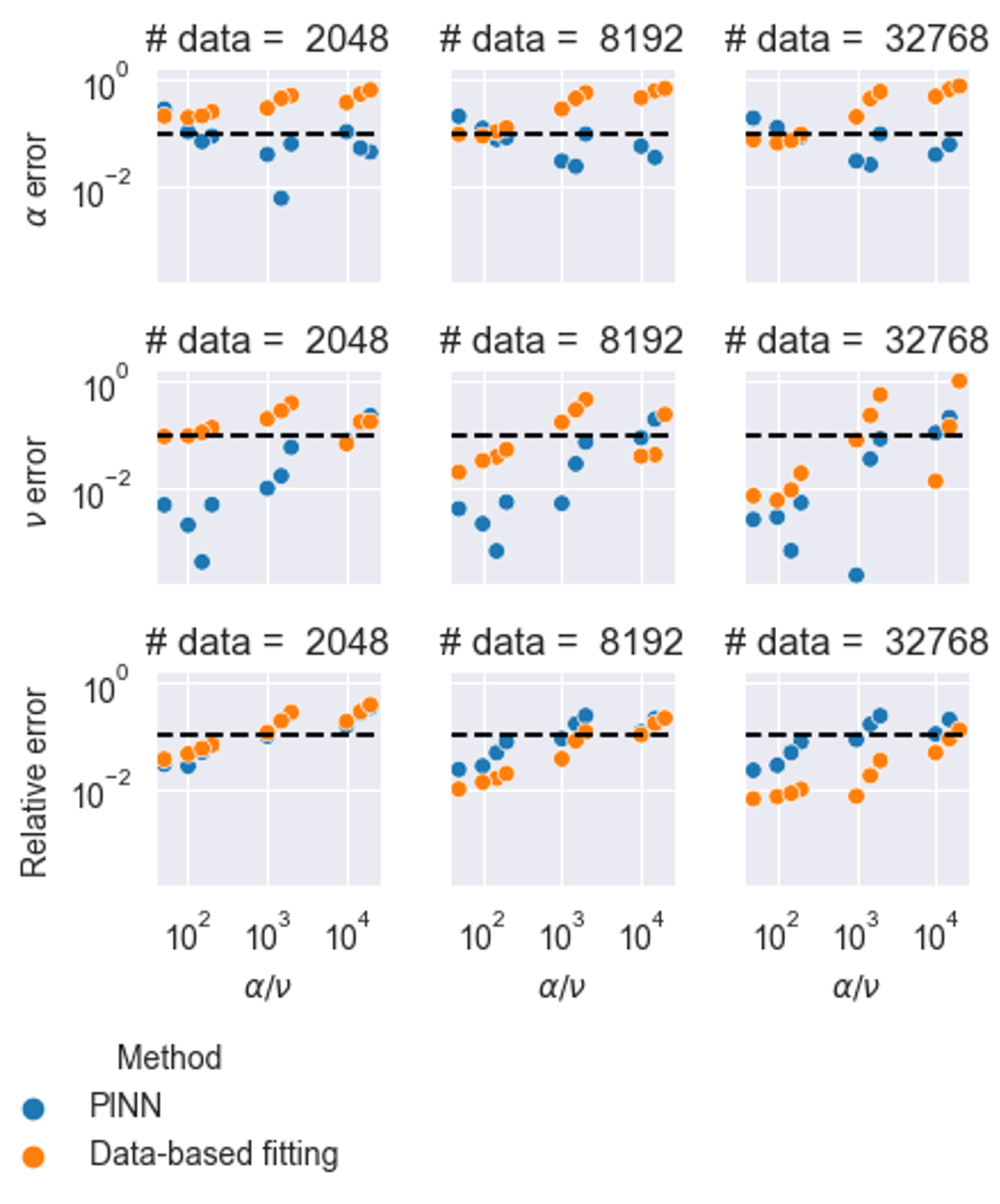}
    \caption{For given quantities of data, we compare the estimation accuracy of \glspl{PINN} (using Newton's method) and the data-driven strategy (\cref{alg:estimation}). Needing to minimize both the \gls{PDE} residual and data losses makes \glspl{PINN} less effective at fitting the solution directly, yielding typically higher relative errors. However, the \glspl{PINN} are generally better at fitting the \gls{PDE} parameters $\alpha$ and $\nu$.}
    \label{fig:comparison}
\end{figure}

In~\cref{fig:comparison}, we summarize the results for \glspl{PINN} and the data-driven strategy in~\cref{fig:comparison}, where $2048, 8192$, or $32768$ training points were supplied.

\glspl{PINN} are trained using a multi-objective loss function based on data and \glspl{PDE}. This can make their convergence noisy and each criteria difficult to simultaneously satisfy. Thus, the data-driven approach can be better at fitting the supplied training data and attaining low relative error on the full solution.

However, fitting only the data to between $1\%$ and $10\%$ relative error generally does not guarantee that the \gls{PDE} parameters are estimated to within $10\%$ error or less. In contrast, \glspl{PINN} are able to more consistently estimate $\nu$ and $\alpha$ given the amount of data provided. \gls{PINN} parameter estimation can be successful with only $8192$ training points, i.e., approximately $0.2\%$ of the solution.

Both methods are less accurate in the regime where $\alpha / \nu$ is large. This is unsurprising, as the fluid flow changes rapidly across temporal and spatial scales in that parameter regime. This echoes work in \gls{PINN} forward modeling highlighting that model predictions can be inaccurate as parameters driving system complexity vary~\cite{krishnapriyan2021characterizing,New2023pinns}.

\subsection{Analysis of results}\label{sec:analysis}

\begin{figure*}[h]
    \centering
    \includegraphics[width=0.6\linewidth]{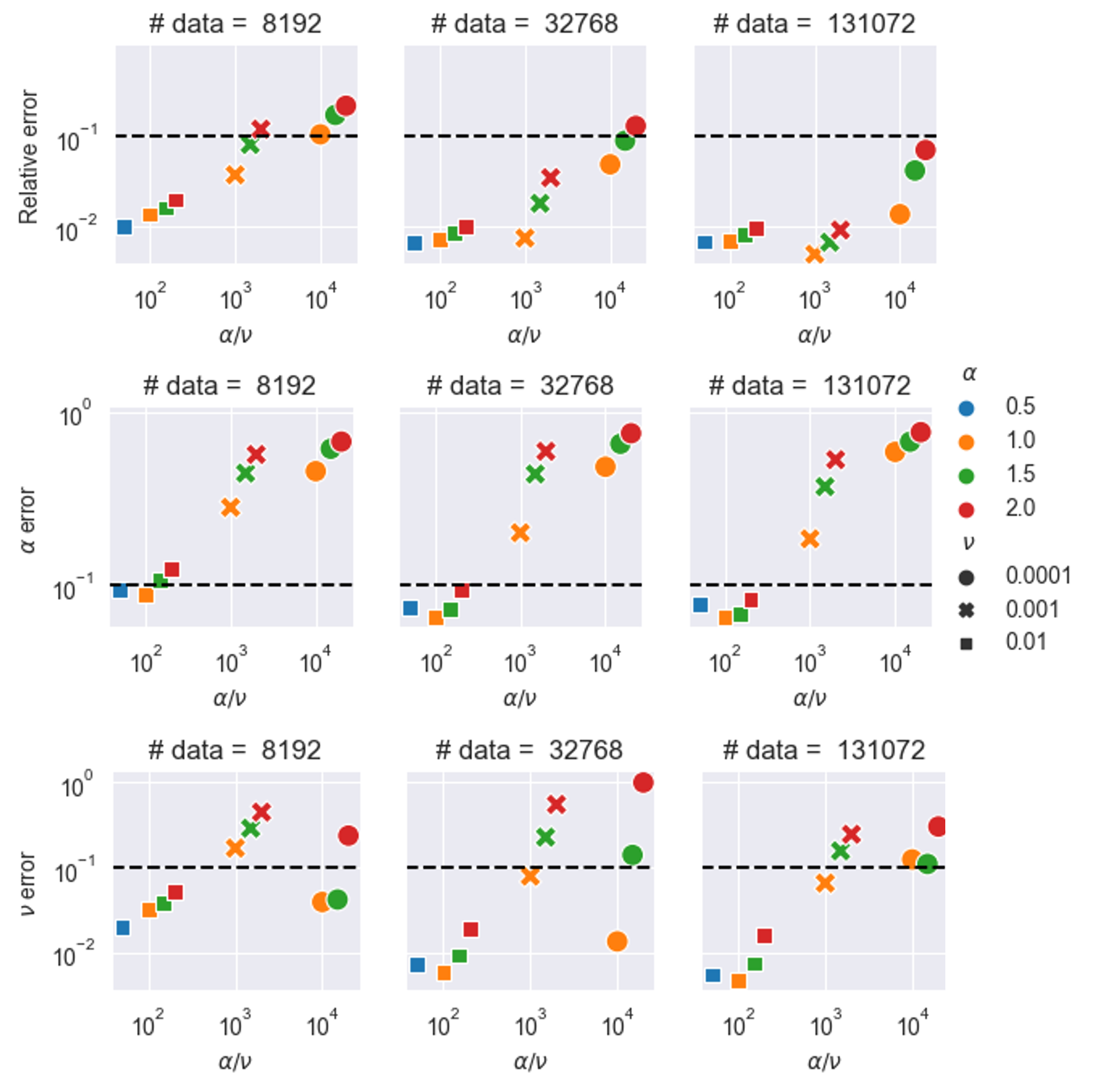}
    \caption{For varying amounts of training data, we plot the relative errors in estimating the Burgers' solution (top), convection coefficient $\alpha$ (middle) and diffusion coefficient $\nu$ (bottom), using the data-driven \gls{NN} (\cref{alg:estimation}) strategy across different Burgers' parameters. The black dashed line indicates $10\%$ or less error, our threshold for success. With $131072$ points, the \gls{NN} can achieve less than $10\%$ solution relative error for every parameter configuration. However, even with $131072$ data points, it struggles to estimate parameters, failing at estimating $\alpha$ for every configuration other than $\nu=0.01$ and failing at estimating $\nu$ in five out of the ten parameter configurations.}
    \label{fig:baseline_results}
\end{figure*}

\begin{figure*}[h]
    \centering
    \includegraphics[width=\linewidth]{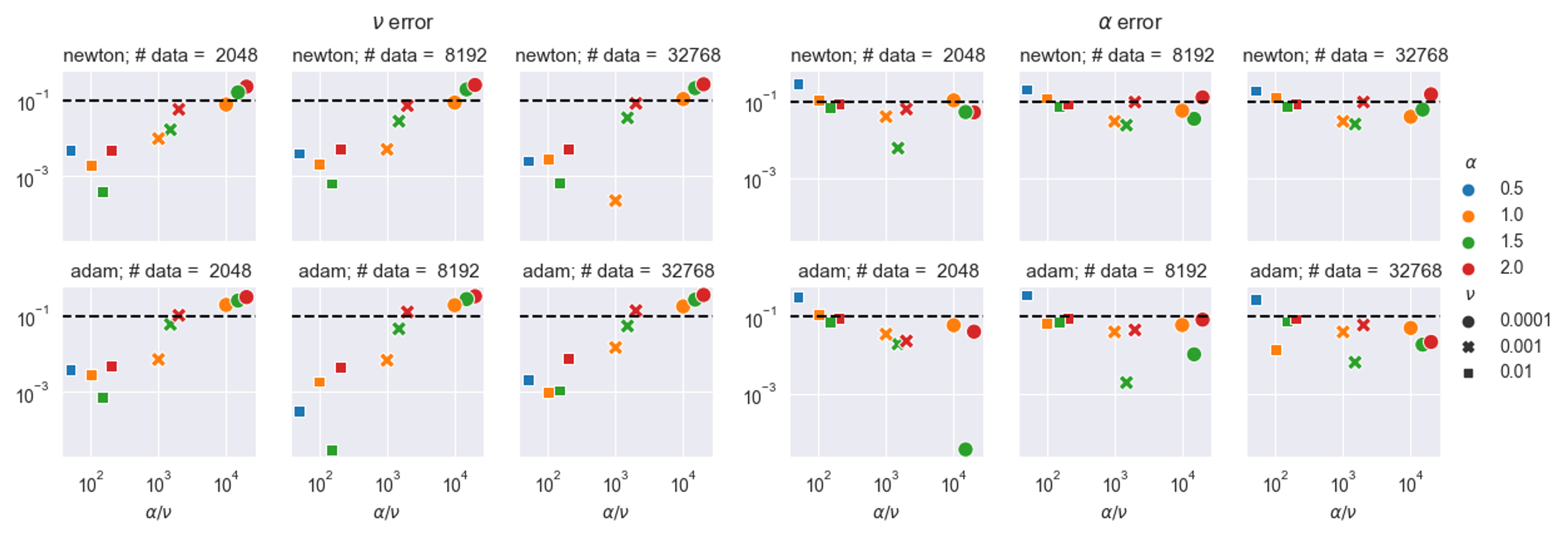}
    \caption{For varying amounts of training data (columns) and different \gls{PDE} optimizers (rows), we plot the relative errors in estimating the diffusion coefficient $\nu$ (left) and convection coefficient $\alpha$ (right), using \glspl{PINN}. The black dashed line indicates $10\%$ or less error, our threshold for success. Compared to the data-driven baseline  (\cref{fig:baseline_results}), \glspl{PINN} are more successful in recovering parameters at a given quantity of labeled training data. They can correctly estimate $\nu$ across values of $\alpha$ except when $\nu=0.0001$ (i.e., when the fluid is highly inviscid). They are also successful in estimating $\alpha$, except in the $\nu=0.01$ regime.}
    \label{fig:pinn_results}
\end{figure*}

\Cref{fig:baseline_results} contains further details about the performance of the data-driven estimation strategy, in the settings where $8192$ or more data points are supplied. Results with fewer training points are in \cref{fig:baseline_error,fig:baseline_alpha}, and \cref{fig:baseline_nu} in~\cref{sec:supp_figures}. In \cref{fig:pinn_results}, as well as in~\cref{fig:pinn_accuracy} in~\cref{sec:supp_figures}, we show further results in terms of parameter estimation and accuracy for the \glspl{PINN}. 

With low amounts of labeled data,~\cref{alg:estimation} unsurprisingly performs poorly, but even in high-data settings (e.g., $131072$ training points, ten times as many as the \glspl{PINN} have), parameter estimation often still fails. We note that the largest number of supplied training points we use, $131072$, is still less than $3\%$ of the total number of points in the solution. Training on larger datasets could enable more reliable parameter estimation.

Although the \glspl{PINN} are more accurate than~\cref{alg:estimation}, similar trends to~\cref{fig:baseline_results} hold, namely that performance degrades in the regime where $\alpha / \nu$ is large. Increasingly, increasing the amount of data supplied does not consistently improve \gls{PINN} accuracy, in comparison to~\cref{alg:estimation}. This is likely a consequence of the underlying complexity and ill-conditionedness of the \gls{PINN} optimization problem.

\begin{table}[!ht]
    \centering
    \begin{adjustbox}{width=0.48\textwidth}
    \begin{tabular}{l|l|l|l|l}
        \gls{PDE} optimizer & \# data & $\alpha$ error & Relative error & $\nu$ error\\ \hline\hline
        Adam & $2048$ & $0.080$ & $0.152$ & $0.093$ \\
        Newton & $2048$ & $0.086$ & $0.150$ & $0.056$ \\\hline
        Adam & $8192$ & $0.085$ & $0.139$ & $0.097$ \\
        Newton & $8192$ & $0.085$ & $0.130$ & $0.065$ \\\hline
        Adam & $32768$ & $0.067$ & $0.139$ & $0.101$ \\
        Newton & $32768$ & $0.088$ & $0.126$ & $0.071$
    \end{tabular}
    \end{adjustbox}
    \caption{We show relative errors and estimation errors for \glspl{PINN}, averaged over the parameter space, comparing using Adam and Newton's method for the \gls{PDE} estimation component. Newton's enables better estimation of $\nu$ and better fitting of the solution, while Adam yields superior estimation of $\alpha$.}
    \label{tab:newton_comparison}
\end{table}

\section{Conclusion}

We have presented a novel \glspl{PINN} benchmark for the vector 2D Burgers' equation with varying equation parameters in the inverse problem setting. Our strategy combines \gls{SGD} and Newton's method to learn the \gls{NN} and \gls{PDE} parameters, respectively. This is a first step towards completion of a major gap in the \glspl{PINN} literature. First we demonstrated the recovery of \gls{PDE} parameters in inverse problems across several viscous and inviscid flow conditions. Second, we addressed the limitations of optimizers such as Adam in estimating \gls{PDE} parameters via the use of Newton's method.

As future work, we advocate for further development of challenging inverse problems, including those that include partially-known \glspl{BC} or \glspl{IC}~\cite{mattey2022novel}, or systematic application of noise  to data solution data (e.g.,~\cite{Hao2023pinnacle}). For these and more challenging settings, we expect that additional training strategies should be employed, such as improved sampling~\cite{wang2024piratenets}. 

In the forward problem \gls{PINN} literature, breaking time domains into smaller subsets and training \glspl{PINN} sequentially on each subset is a common strategy~\cite{krishnapriyan2021characterizing,Wang2022causality}, and predicting phenomena across large time domains is a known challenge for \glspl{PINN}~\cite{Meng2020parareal,Wang2022causality,Daw2022mitigating}.
Here, we estimated parameters from solutions confined to $t\in[0, 0.499]$, but using either a smaller or large time domain could have enabled generally better parameter estimation. Thus, we suggest deeper exploration into the impact of time domain size on the parameter estimation problem, especially when the data feature discontinuities or sudden changes.


\section*{Acknowledgments}
This work was supported by internal research and development funding from the Research and Exploratory Development Mission Area of the Johns Hopkins University Applied Physics Laboratory.



\bibliography{references}
\bibliographystyle{icml2024}

\clearpage
\newpage
\onecolumn
\appendix
\section{Hyperparameters}\label{sec:hyperparameters}

\begin{table}[h!]
    \centering
    \begin{tabular}{c|c}
    Hyperparameter              &   Value\\\hline\hline
    Number of hidden units      &   $256$\\
    Number of layers            &   $10$\\
    Activation function         &   $\tanh$\\\hline
    Batch size (Data loss)      &   $2048$\\
    Batch size (\gls{IC} loss)  &   $2048$\\
    Batch size (\gls{PDE} loss) &   $8192$\\\hline
    $\lambda_{Data}$            &   1\\
    $\lambda_{IC}$              &   1\\\hline
    Optimizer                   &   Adam\\
    Number of epochs            &   $50000$\\
    Initial learning rate       &   $5\cdot10^{-3}$\\
    Minimum learning rate       &   $10^{-6}$\\
    Exponential decay rate      &   $0.925$\\
    Exponential decay interval  &   $5000$\\
    \end{tabular}
    \caption{Hyperparameters used for the data-driven estimation strategy (\cref{alg:estimation}). For the data loss, we use a batch size of the minimum of the number of labeled data points and $2048$. Models use the \gls{EDMLP}~\cite{wang2021understanding}, modified to exactly satisfy the \glspl{BC} (eqs.~\ref{eq:exact}).}
    \label{tab:baseline_parameters}
\end{table}

\begin{table}[h!]
    \centering
    \begin{tabular}{c|c}
    Hyperparameter                  &   Value\\\hline\hline
    Number of hidden units          &   $256$\\
    Number of layers                &   $10$\\
    Activation function             &   $\tanh$\\\hline
    Batch size (Data loss)          &   Varies\\
    Batch size (\gls{IC} loss)      &   $1024$\\
    Batch size (\gls{PDE} loss)     &   $2048$\\\hline
    $\lambda_{PDE}$                 &   $1$\\
    $\lambda_{Data}$                &   $10$\\
    $\lambda_{IC}$                  &   $10$\\\hline
    \gls{NN} Optimizer              &   Adam\\
    \gls{PDE} Optimizer             &   \{Adam, Newton\}\\
    Number of gradient steps        &   $100000$\\
    Parameter estimation interval   &   $\{10\,(\text{Adam}),100\,(\text{Newton})\}$\\
    Initial \gls{NN}  learning rate &   $5\cdot10^{-3}$\\
    Minimum \gls{NN} learning rate  &   $10^{-5}$\\
    Exponential decay rate          &   $0.925$\\
    Exponential decay interval      &   $5000$\\
    \gls{PDE} parameter step size   &   $10^{-3}$
    \end{tabular}
    \caption{Hyperparameters used for \glspl{PINN}. For the data loss, we use all available labeled data points for each gradient update. Models use the \gls{EDMLP}~\cite{wang2021understanding}, modified to exactly satisfy the \glspl{BC} (eqs.~\ref{eq:exact}). 
    When training, we only update the \gls{PDE} parameters every $n$ epochs, where $n=10$ if the \gls{PDE} optimizer is Adam, and $n=100$ if the \gls{PDE} parameter optimizer is Newton's method.}
    \label{tab:pinn_parameters}
\end{table}

\onecolumn

\section{Supplementary Figures}\label{sec:supp_figures}

\begin{figure*}[h]
    \centering
    \includegraphics[width=0.7\linewidth]{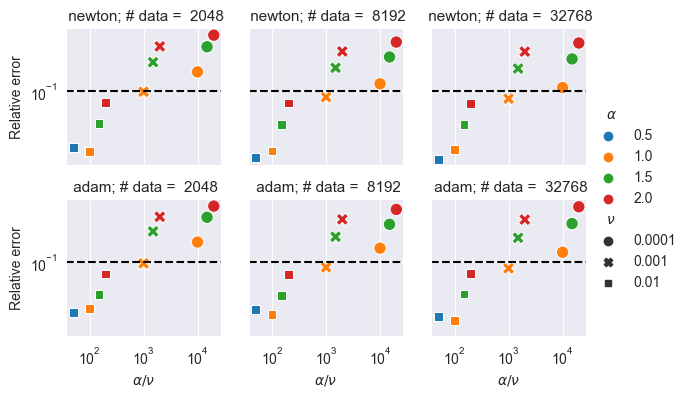}
    \caption{For varying amounts of training data (columns) and different \gls{PDE} optimizers (rows), we plot the relative errors in the predicted solutions to the Burgers' equation.}
    \label{fig:pinn_accuracy}
\end{figure*}

\begin{figure}[h]
    \centering
    \includegraphics[width=0.7\linewidth]{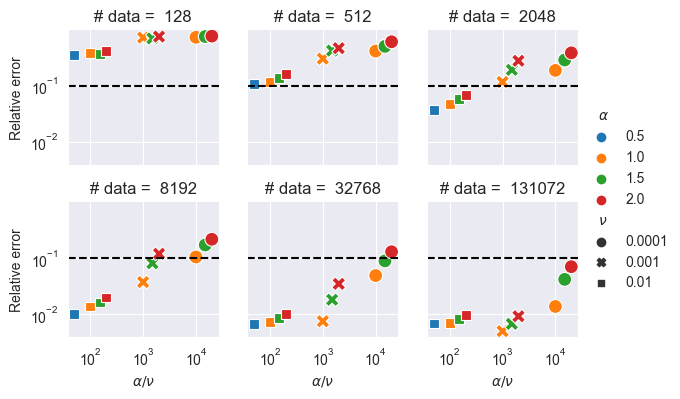}
    \caption{For varying amounts of training data, we plot the relative errors in estimating the Burgers' solution using the data-driven \gls{NN} (\cref{alg:estimation}) strategy across different Burgers' parameters. The black dashed line indicates $10\%$ or less error, our threshold for success. }
    \label{fig:baseline_error}
\end{figure}

\begin{figure}[h]
    \centering
    \includegraphics[width=0.7\linewidth]{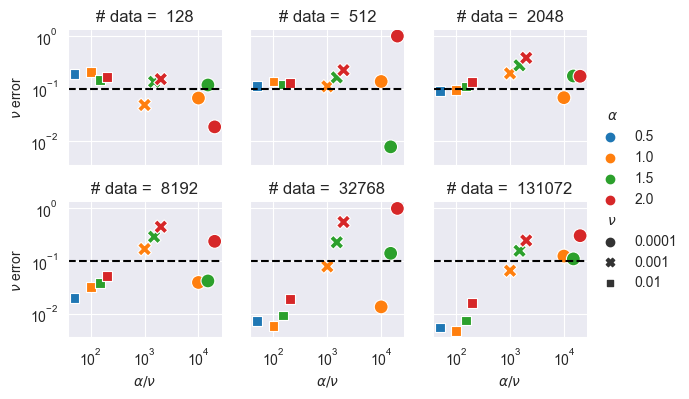}
    \caption{For varying amounts of training data, we plot the relative errors in estimating the convection coefficient $\alpha$ using the data-driven \gls{NN} (\cref{alg:estimation}) strategy across different Burgers' parameters. The black dashed line indicates $10\%$ or less error, our threshold for success. }
    \label{fig:baseline_alpha}
\end{figure}

\begin{figure}[h]
    \centering
    \includegraphics[width=0.7\linewidth]{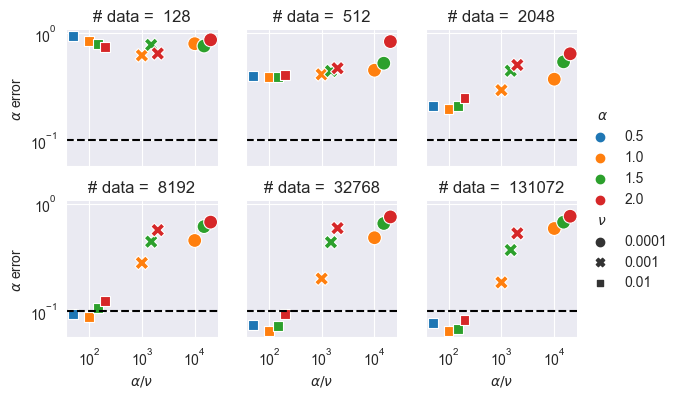}
    \caption{For varying amounts of training data, we plot the relative errors in estimating the diffusion coefficient $\nu$, using the data-driven \gls{NN} (\cref{alg:estimation}) strategy across different Burgers' parameters. The black dashed line indicates $10\%$ or less error, our threshold for success.}
    \label{fig:baseline_nu}
\end{figure}


\end{document}

%% file: math_commands.tex
\usepackage{amsmath,amsfonts,bm}

\def\eqref#1{equation~\ref{#1}}

\def\1{\bm{1}}

\def\rvu{{\mathbf{i}}}

\def\rvu{{\mathbf{u}}}

\DeclareMathAlphabet{\mathsfit}{\encodingdefault}{\sfdefault}{m}{sl}
\SetMathAlphabet{\mathsfit}{bold}{\encodingdefault}{\sfdefault}{bx}{n}



%% file: glossary.tex
\newacronym{APL}{APL}{%
  the Johns Hopkins University Applied Physics Laboratory%
}
\newacronym{AD}{AD}{automatic differentiation}

\newacronym{BC}{BC}{boundary condition}

\newacronym{CFD}{CFD}{computational fluid dynamics}
\newacronym{CSPINN}{CS-PINN}{constraint-satisfaction physics-informed neural network}

\newacronym{DNS}{DNS}{direct numerical simulation}
\newacronym{DE}{DE}{differential equation}
\newacronym{EDMLP}{ED-MLP}{encoder-decoder multi-layer perceptron}

\newacronym{FFD}{FFD}{fast fluid dynamics}

\newacronym{FD}{FD}{finite difference}
\newacronym{FEM}{FEM}{finite element method}
\newacronym{FVM}{FVM}{finite volume method}
\newacronym{FMLP}{F-MLP}{Fourier embedding multi-layer perceptron}

\newacronym[plural=GPUs,longplural={graphics processing units}]{GPU}{GPU}{%
  graphics processing unit%
}
\newacronym{GD}{GD}{gradient descent}
\newacronym{GAAF}{GAAF}{globally adaptive activation function}


\newacronym{IC}{IC}{initial condition}

\newacronym{JHU}{JHU}{%
  Johns Hopkins University%
}
\newacronym{JIT}{JIT}{just-in-time}


\newacronym{LAAF}{LAAF}{locally adaptive activation function}

\newacronym{ML}{ML}{machine learning}
\newacronym{MLP}{MLP}{multi-layer perceptron}

\newacronym{NN}{NN}{neural network}
\newacronym{NTK}{NTK}{neural tangent kernel}

\newacronym{ODE}{ODE}{ordinary differential equation}

\newacronym[
    shortplural={PINNs}
]{PINN}{PINN}{physics-informed neural network}
\newacronym{PDE}{PDE}{partial differential equation}
\newacronym{PMLP}{P-MLP}{perioidic multi-layer perceptron}

\newacronym{ROM}{ROM}{reduced order model}

\newacronym{SciML}{SciML}{scientific machine learning}
\newacronym{SGD}{SGD}{stochastic gradient descent}


